\documentclass[12pt]{article}

\usepackage{amsmath,amsgen,amstext,amsbsy,amsopn,amsthm, amssymb}

\newtheorem{thm}{Theorem}
\numberwithin{thm}{section}

\numberwithin{cor}{section}
\newtheorem{lem}{Lemma}
\numberwithin{lem}{section}

\numberwithin{prop}{section}
\theoremstyle{definition}

\numberwithin{defn}{section}

\numberwithin{rem}{section}

\newcommand{\suli}{\sum\limits}

\newcommand{\al}{\alpha}

\newcommand{\E}{\mathcal{E}}

\newcommand{\rmax}{\rho_{\al,\text{max}}}
\newcommand{\rmin}{\rho_{\al,\text{min}}}
\newcommand{\R}{\mathbb{R}}

\numberwithin{equation}{section}

\def\p{{\vec p}}

\def\R{{\bf R}}

\def\E{{\cal E}}

\def\x{{\vec{x}}}

\def\mfr#1/#2{\hbox{$\frac{{#1} }{ {#2}}$}}

\normalbaselines 
\begin{document}

\begin{center}
\vspace*{1.0cm}

{\LARGE{\bf The Ground State Energy and Density \\
of Interacting Bosons in a Trap}} 

\vskip 1.5cm

{\large {\bf Elliott~H.~Lieb$^{1}$, Robert Seiringer$^2$ and Jakob
Yngvason$^2$}}

\vskip 0.5cm

$^1$Departments of Physics and Mathematics, Princeton University, \\
P.~O.~Box 708, Princeton, New Jersey 08544-0708, USA\\ 
$^2$ Institut f\"ur Theoretische Physik, Universtit\"at Wien, \\
Boltzmanngasse 5, A-1090 Wien, Austria

\end{center}

\vspace{1 cm}

\begin{abstract}
In the theoretical description of recent experiments with dilute Bose 
gases confined in external potentials the 
Gross-Pitaevskii equation plays an important role. Its status as 
an approximation for the quantum mechanical many-body ground state
problem has recently been rigorously clarified. A summary of this work is 
presented here.
\end{abstract}

\footnotetext[0]{\copyright 1999 by the authors. Reproduction
of this work, in its entirety, by any means, is permitted for
non-commercial purposes.}

\vspace{1 cm} 

\section{Introduction}

The Gross-Pitaevskii (GP) equation is a nonlinear Schr\"odinger equation 
that was introduced in the early sixties  \cite{G1961}--\cite{G1963}  as a 
phenomenological equation for the order parameter in superfluid 
${\rm He}_{4}$. It has come into prominence again because of recent 
experiments on Bose-Einstein condensation of dilute gases in magnetic traps.
The paper \cite{DGPS} brings an up to date review of these 
developments.

One of the inputs needed for the justification of the GP equation 
starting from the many body Hamiltonian is the ground state energy of 
a a dilute, thermodynamically infinite, homogeneous Bose gas.  The 
formula for this quantity is older than the GP equation but it has 
only very recently been derived rigorously for suitable interparticle 
potentials.  See \cite{LY1998} and \cite{LY1999}.  The paper 
\cite{LSY1999} goes one step further and derives the GP as 
a limit of the full quantum mechanical description. This derivation 
is summarized in the present contribution.

The starting point of the investigation is the 
Hamiltonian  for $N$ bosons that interact with each other via a 
spherically symmetric pair-potential $v(|\x_i - \x_j|)$ and are 
confined by an external potential $V(\x)$:
\begin{equation}
H = \sum_{i=1}^{N} \{- \Delta_i + V(\x_{i})\}+
\sum_{1 \leq i < j \leq N} v(|\x_i - \x_j|).
\end{equation}
The Hamiltonian acts on {\it symmetric} wave functions in 
$L^2(\R^{3N},d^{3N}x)$.  The pair interaction $v$ is assumed to be 
{\it nonnegative} and of short range, more precisely, $v(r)\leq {\rm 
(const.)}\,r^{-(3+\varepsilon)}$ as $r\to\infty$, for some 
$\varepsilon>0$.  The potential $V$ that represents the trap is 
continuous and $V(\x)\to\infty$ as $|\x|\to\infty$.  By shifting the 
energy scale we can assume that $\min_{\x}V(\x)=0$.

Units are chosen so that $\hbar=2m=1$, where $m$ is the 
particle mass. A natural energy unit is given by the ground state
energy $\hbar\omega$ of the one particle Hamiltonian $-(\hbar^2/2m)\Delta+V$.
The corresponding length unit, $\sqrt{\hbar/(m\omega)}$, measures the 
effective extension of the trap.

The ground state wave function $\Psi_{0}(\x_{1},\dots,\x_{N})$ 
satisfies 
$H\Psi_{0}=E^{\rm QM}\Psi_{0}$, 
with the ground state energy $E^{\rm QM}=\inf{\rm \,spec\,}H$.
If $v=0$, then 
$\Psi_{0}(\x_{1},\dots,\x_{N})=
	\prod_{i=1}^{N}\Phi_{0}(\x_{i}),$
with $\Phi_{0}$ the ground state wave function of $- \Delta + V(\x)$.
On the other hand, if
$v\neq 0$ the ground state 
$\Psi_{0}(\x_{1},\dots,\x_{N})$ is, in general,
far from being a product state if 
$N$ is large. In spite 
of this fact,  recent experiments on BE condensation are usually 
interpreted in terms of a function $\Phi^{\rm GP}(\x)$ of a single 
$\x\in{\mathbb 
R}^3$, the solution of the
{\it Gross-Pitaevskii equation}
\begin{equation}\label{gpe}(- \Delta + V+8\pi 
a|\Phi|^2)\Phi=\lambda\Phi,   \end{equation}
together with the normalization condition
\begin{equation}\label{norm}
	\int_{{\mathbb R}^3}|\Phi(\x)|^2=N.
\end{equation}
Here $a$ is the {\it scattering length} of the potential $v$:
\begin{equation}a=\lim_{r\to\infty}\left(r-\frac{u_{0}(r)}{ 
u_{0}'(r)}\right)\end{equation}
where $u_{0}$ satisfies the zero energy scattering equation,
\begin{equation}\label{scatteq}- u^{\prime\prime}(r)+\mfr1/2 v(r) 
u(r)=0,\end{equation}
and $u_{0}(0)=0$. (The factor $1/2$ is due to the reduced mass of the 
two-body problem.)
The GP equation (\ref{gpe}) is the variational equation for the 
minimization of the GP {\it energy 
functional}
\begin{equation}\label{gpf}\E^{\rm 
GP}[\Phi]=\int\left(|\nabla\Phi|^2+V|\Phi|^2+4\pi 
a|\Phi|^4\right)d^3\x\end{equation}
with the subsidiary condition (\ref{norm}). The corresponding energy is
\begin{equation}E^{\rm GP}(N,a)=\inf_{\int|\Phi|^2=N}\E^{\rm GP}[\Phi]=
	\E^{\rm GP}[\Phi^{\rm GP}
],\end{equation}
with a unique, positive $\Phi^{\rm GP}$. The eigenvalue,
$\lambda$,  in (\ref{gpe}) is related to $E^{\rm GP}$ by
\begin{equation}
\lambda=dE^{\rm GP}(N,a)/dN=E^{\rm GP}(N,a)/N+4\pi a\bar \rho,	
	\end{equation}
	where
	\begin{equation}\label{rhobar}
\bar\rho=\frac 1N\int|\Phi^{\rm GP}(\x)|^4 d^3\x
\end{equation}
is the {\it mean density}. The minimizer $\Phi^{\rm GP}$ of
(\ref{gpf}) with the condition (\ref{norm}) depends on $N$ and $a$, of
course, and when this is important we denote it by  $\Phi^{\rm GP}_{N,a}$.

Mathematically, the GP equation is quite similar to the
Thomas-Fermi-von Weizs\-\"acker equation \cite{lieb81} and its basic
properties can be established by similar means. See \cite{LSY1999},
Sect.~2 and Appendix A.

The idea is now that for {\it dilute} gases one should have
\begin{equation}\label{approx}E^{\rm GP}
	\approx E^{\rm QM}\quad{\rm and}\quad \rho^{\rm QM}(\x)\approx
\left|\Phi^{\rm GP}(\x)\right|^2\equiv \rho^{\rm GP}(\x),\end{equation}
where the quantum mechanical particle density in the ground state is
defined by \begin{equation} \rho^{\rm
QM}(\x)=N\int|\Psi_{0}(\x,\x_{2},\dots,\x_{N})|^2d\x_{2}\cdots
d\x_{N}.  \end{equation} {\it Dilute}  means here that
\begin{equation}\bar\rho a^3\ll 1.\end{equation} The task is to make
(\ref{approx}) precise and prove it!

The first remark is that by scaling
\begin{equation}E^{\rm GP}(N,a)=NE^{\rm GP}(1,Na)\quad
\mbox{and}\quad
\Phi^{\rm GP}_{N,a}=N^{1/2}\Phi^{\rm GP}_{1,Na}.
\label{scaling}
\end{equation}
Hence $Na$ is the natural parameter in GP theory.  It should also be 
noted that $E^{\rm QM}$ depends on $N$ and $v$ and not only on the 
scattering length $a$.  However, in the limit we are about to define, 
it is really only $a$ that matters. To bring this out we write
\begin{equation}v(r)=(a_1/a)^2v_1(a_1r/a),
\end{equation}
where $v_{1}$ has scattering length $a_{1}$ and regard $v_{1}$ as {\it 
fixed}. Then $E^{\rm QM}$ is a function $E^{\rm QM}(N,a)$ 
of $N$ and $a$ and we can 
state our main result.

\begin{thm}[Dilute limit of the QM ground state energy and density]
\begin{equation}\label{econv}
	\lim_{N\to\infty}\frac{{E^{\rm QM}(N,a)}}{ {E^{\rm 
GP}(N,a)}}=1\end{equation}
and
\begin{equation}\label{dconv}
	\lim_{N\to\infty}\frac{1}{ N}\rho^{\rm QM}_{N,a}(\x)=
\left |{\Phi^{\rm GP}_{1,Na}}(\x)\right|^2\end{equation}
 if $Na$ is fixed. The convergence in (\ref{dconv}) is in the 
 weak $L_1$-sense.
 \end{thm}
\noindent In particular, the limits depend only on the scattering
length $a_{1}=Na$ of $v_{1}$ and not on details of the potential.
\medskip

 \noindent{\it Remark.} Since $\bar\rho\sim N$ the attribute `dilute' 
 may seem a bit strange for these limits.  However, what matters is 
 that the scattering length $a$ is small compared to the mean particle 
 distance, $\sim\bar\rho^{1/3}$, and if $Na$ is kept constant, then 
 $\bar\rho a^3\sim N^{-2}$.  It is also important to remember that the 
 unit of length, $\sqrt{\hbar/(m\omega)}\equiv a_{\rm trap}$, depends on 
 the external potential, and $a$ really stands for $a/a_{\rm trap}$.  
 If the external potential is scaled with $N$, the both $a_{\rm trap}$ 
 and the energy unit $\hbar\omega$ depend on $N$.  For instance, in 
 order to achieve a finite transition temperature for the BE 
 condensation of a noninteracting gas in a parabolic trap it is 
 necessary to keep $N\omega^3$ fixed in the thermodynamic limit, and 
 hence $a_{\rm trap}\sim N^{1/6}$ and $\hbar\omega\sim N^{-1/3}$.  
 (See \cite{DGPS}, Eq.\ (14).)  However, since the energy unit cancels in 
 (\ref{econv}) a dependence of $V$ on $N$ does not affect the 
 validity of (\ref{econv}), and (\ref{dconv}) also remains valid taking
 into account that both sides really contain the factor 
 $a_{\rm trap}^{-3}$.
 
 \section{The dilute homogeneous Bose gas}

The motivation for the last term in the GP energy functional
(\ref{gpf}) is an asymptotic formula for the quantum mechanical ground
state energy $E_{0}(N,L)$ of $N$ bosons in a  rectangular {\it box} of
side length $L$ (i.e., the {\it homogeneous} case), that was put
forward by several authors many decades ago.  Consider the energy per
particle in the
thermodynamic limit with $\rho=N/L^3$ fixed:
\begin{equation}e_{0}(\rho)=
	\lim_{L\to\infty}E_{0}(\rho L^3,L)/(\rho L^3).\end{equation}
According to the pioneering work of Bogoliubov \cite{BO} the leading 
term in the
{\it low density asymptotics} of $e_{0}(\rho)$ is given by
\begin{equation}e_{0}(\rho)\approx4\pi\rho a\end{equation}
for $\rho a^3\ll 1$.
In the 50's and early 60's several derivations of this formula were 
presented \cite{Lee-Huang-YangEtc}.
They all depended on some special 
assumptions about the ground state that have never been proved or the 
selection of some special terms from a perturbation expansion that 
most likely diverges.
The only rigorous estimates of this period were obtained by Dyson \cite{dyson}
for hard spheres:
\begin{equation}\frac{1+2 Y^{1/3}}{  (1-Y^{1/3})^2}\geq 
\frac{e_{0}(\rho)}{ 4\pi\rho a}\geq\frac{1}{ 10\sqrt 2} \end{equation}
with $Y\equiv 4\pi\rho a^3/3$. The upper bound has the right 
asymptotic form, and it is not very difficult to generalize it to other 
potentials than hard spheres.
The lower bound on the other hand  is too low by a factor of about $1/14$
and a remedy for this situation was obtained only 40 years 
after Dyson's paper:
\begin{thm}[Lower bound for a homogeneous gas] If $v$ is nonnegative and of 
finite range, then
\begin{equation}\label{lbd}\frac{e_{0}(\rho)}{ 4\pi\rho a}\geq (1-C\, 
Y^{1/17})\end{equation}
\medskip
with some constant $C$.
\end{thm}
The proof of this theorem is given in  \cite{LY1998}, see also the 
exposition in \cite{LY1999}. For the application to the proof of Theorem 1.1 
we need a version for finite boxes that is implicitly contained in 
\cite{LY1998} and explicitly stated in \cite{LY1999}:
\begin{thm}[Lower bound in a finite box] \label{lbthm2} 
	For a  positive potential $v$ with finite range there is 
a $\delta>0$ such that 
\begin{equation}\label{lbd2}E_{0}(N,L)/N\geq 4\pi\mu\rho 
a \left(1-C\, 
Y^{1/17}\right)
\end{equation} 
for all $N$ and $L$ with $Y<\delta$ and $L/a>C'Y^{-6/17}$. Here 
$C$ and $C'$ are constants,
independent of $N$ and $L$. (Note that the condition on $L/a$
 requires in particular that $N$ must be large enough, 
 $N>\hbox{\rm (const.)}Y^{-1/17}$.) 
 \end{thm}
These results are stated for interactions of finite range. An
extension to potentials $v$ of infinite range 
decreasing faster than $1/r^3$ at infinity is 
obtained by approximating $v$ by finite range potentials, controlling the 
change of the scattering length as the cut-off is removed. In this 
case the estimate holds also, but possibly with an exponent different 
from $1/17$ and a different constant.
It should be noted, however, that the form of the error term in 
(\ref{lbd}) is 
dictated by the method of proof. The true error term 
presumably
does not have a negative sign for sufficiently small $Y$. 

\section{An upper bound for the QM energy}

We now turn to the inhomogeneous gas. In order to prove Eq.\ (\ref{econv}) one 
has to establish upper and 
lower bounds for $E^{\rm QM}$ in terms of $E^{\rm GP}$ with errors 
that vanish in the limit considered. As usual, the upper bound 
is easier. It is based on test wave 
functions of the form
\begin{equation}\label{ansatz}\Psi(\x_{1},\dots,\x_{N})
	=\prod_{i=1}^N\Phi^{\rm 
GP}(\x_{i})F(\x_{1},\dots,\x_{N}).\end{equation}
where $F$ is constructed in the following way:
\begin{equation}F(\x_1,\dots,\x_N)=\prod_{i=1}^N 
f(t_i(\x_1,\dots,\x_i)),\end{equation}
where $t_i = \min\{|\x_i-\x_j|, 1\leq j\leq i-1\}$ is the distance 
of $\x_{i}$ to its {\it nearest neighbor} among the points 
$\x_1,\dots,\x_{i-1}$ and $f$ is a  function of $t\geq 0$.
With $u_{0}$ the zero energy scattering solution and 
$f_{0}(r)=u_{0}(r)/r$ the function $f$ can be taken as
\begin{equation}\label{eff}f(r)=f_{0}(r)/f_{0}(b)\end{equation}
for $r<b\equiv (4\pi\bar\rho/3)^{-1/3}$ and 1 otherwise. The function 
(\ref{ansatz}) is not totally symmetric, but for an upper bound it is 
nevertheless an acceptable test wave function 
since the bosonic ground state energy is equal to the 
{\it absolute} ground state energy.
 
The result of a somewhat lengthy computation is the upper bound
\begin{equation}\label{ubd}E^{\rm QM}\leq E^{\rm GP}(1+O(\bar Y^{1/3}))\end{equation}
with $\bar Y=4\pi a^3\bar\rho/3$. Note that $\bar Y\sim N^{-2}$ 
since $a\sim N^{-1}$ and
$\bar\rho\sim N$.

\section{The lower bound} 

To obtain a lower bound for the QM energy in an external potential the 
strategy is to divide space into boxes and use the estimate (\ref{lbd2})
for a homogeneous gas in each box with {\it Neumann} boundary 
conditions. One then minimizes over all 
possible divisions of the particles among the different boxes,
This gives a lower bound to the energy because 
discontinuous wave functions for the quadratic form defined by the 
Hamiltonian are now allowed. Finally, one lets the box size tend to zero. 
 However, it is not possible to simply approximate $V$ by a 
constant potential in each box. To see this consider the case of 
noninteracting particles, i.e., $v=0$ and hence $a=0$. Here
$E^{\rm QM}=N\hbar\omega$, but a `naive' box
method gives only 0 as lower bound, since it clearly 
pays to put all the particles with a constant wave function in the box 
with the lowest value of $V$.

For this reason we start by separating out the GP wave 
function in each variable and write a general wave function $\Psi$ as 
\begin{equation}\Psi(\x_{1},\dots,\x_{N})
	=\prod_{i=1}^N\Phi^{\rm 
GP}(\x_{i})F(\x_{1},\dots,\x_{N})\end{equation}
	This defines $F$ for a given $\Psi$ because $\Phi^{\rm GP}$ is 
everywhere 
	strictly positive, being the ground state of the operator
	$- \Delta + V+8\pi a|\Phi^{\rm GP}|^2$.
	We now compute the expectation value of $H$ in the state $\Psi$, using 
partial integration and  the variational equation (\ref{gpe}) for 
	$\Phi^{\rm GP}$. 
The result is
 \begin{equation}\label{ener2}
\frac{\langle\Psi,H\Psi\rangle}{\langle\Psi,\Psi\rangle}-E^{\rm GP}=4\pi 
a
\bar\rho N+Q(F)
\end{equation}
with
\begin{equation}\label{ener3}
Q(F)=\suli_{i=1}^{N}
\frac{\int\prod_{k=1}^{N}\rho^{\rm GP}(\x_k) 
\left(|\nabla_i 
F|^2+\suli_{j=1}^{i-1}
v(|\x_i-\x_j|)|F|^2-8\pi a\rho^{\rm GP}(\x_i)|F|^2\right)}
{\int\prod_{k=1}^{N}\rho^{\rm GP}(\x_k)|F|^2}.
\end{equation}
Compared to the expression for the energy involving $\Psi$ itself we have thus 
obtained the following replacements:
\begin{equation} V(\x)\rightarrow
 -8\pi a\rho^{\rm GP}(\x),\quad\mbox{\rm and}\quad\prod_{i=1}^Nd\x_i
 \rightarrow \prod_{i=1}^N\rho^{\rm GP}(\x_{i})d\x_{i}.  \end{equation}
(Recall that $\rho^{\rm GP}(\x)=|\Phi^{\rm GP}(\x)|^2$.)
We have to show that the normalized quadratic form $Q$ is bounded below by
$-4\pi a\bar\rho N$, up to small errors, and we use the box method on
{\it this} problem.

Labeling the boxes by an index $\alpha$ we have
\begin{equation}
\inf_F Q(F)\geq \inf_{\{n_\al\}} \suli_\al \inf_{F_\al}Q_\al (F_\al),
\end{equation}
where $Q_\al$ is defined by the same formula as $Q$  but with the 
integrations limited to the box $\alpha$,  $F_{\alpha}$ is a wave function
with particle number $n_\alpha$, and the infimum is taken over all distributions 
of the particles with $\sum n_\al=N$.
We now want to use \eqref{lbd2} and therefore we must approximate  
$\rho^{\rm GP}$ by a  constant in each box. Let 
$\rmax$ and
$\rmin$, respectively, denote the maximal and minimal values of 
$\rho^{\rm GP}$ in box $\al$. We obtain
\begin{equation}\label{qa}
Q_\al(F_{\al})\geq \frac{\rmin}{\rmax}E_0(n_\al,L)-8\pi a\rmax n_\al,
\end{equation}
where $L$ is the side length of each box and $E_{0}(n,L)$ is the 
ground state energy of $n$ bosons in a box without external potential.  
Now by Theorem 1.1 $E_{0}(n_{\al},L)\approx 4\pi a n_{\al}^2/L^3$, 
and 
\begin{equation}\label{estimate}
\inf_{\{n_\al\}}\sum_{\al}(4\pi  n_{\al}^2/L^3-8\pi \rmax n_\al)\geq
-4\pi\sum_{\al}\rmax^2 L^3\approx -4\pi\int\left(\rho^{\rm GP}\right)^2=- 
4\pi\bar\rho N,	
\end{equation}	
which looks promising, were it not for the factor ${\rmin}/{\rmax}$ in
(\ref{qa}) which we would like to replace by 1 for small boxes.  The
problem is that for any fixed size of boxes ${\rmin}/{\rmax}$ tends
rapidly to zero for boxes far from the origin.

This problem can be solved by enclosing the whole system in a big box
$\Lambda_{R}$ of side length $R$.  If $\Phi^{\rm GP}_{R}$ is the
solution of the GP equation with Neumann conditions on the boundary of
$\Lambda_{R}$ (this is the minimizer of a GP functional where the
integration is restricted to $\Lambda_{R}$), then $\rho^{\rm
GP}_{R}=|\Phi^{\rm GP}_{R}|^2$ is bounded from below away from zero in
$\Lambda_{R}$.  Replacing in (\ref{ener3}) everywhere $\rho^{\rm GP}$
by $\rho^{\rm GP}_{R}$ and restricting the integrations to
$\Lambda_{R}$ we can let the side lengths of the small boxes tend to
zero and be sure that ${\rmin}/{\rmax}\to 1$ uniformly for all the
small boxes.

For this method to work, however, we must control the error made by 
enclosing the system in the big box. Let $E^{\rm QM}_{R}(N,a)$ denote the 
quantum mechanical ground state energy in $\Lambda_{R}$ with Neumann 
conditions on the boundary. The essential step is
\begin{lem}
There is an $R_{0}<\infty$, depending only on $Na$ such that
\begin{equation}\label{err}
	E^{\rm QM}(N,a)\geq E^{\rm QM}_{R}(N,a)
\end{equation}	
for all $R\geq R_{0}$ and all $N$, $a$ with $Na$ fixed.
\end{lem}
This lemma follows from $V(\x)\to\infty$ for $|\x|\to\infty$,
together with an
estimate for the chemical potential from above:
\begin{equation}
	E^{\rm QM}_{R}(N+1,a)-E^{\rm QM}_{R}(N,a)\leq e(Na)
(1+O(\bar Y^{1/3}))
\label{49}
\end{equation}
where $e(Na)$ depends only on $Na$ and is independent of $R$. The proof
of (\ref{49}) is similar to the proof of the upper bound (\ref{ubd})
but this time one uses
\begin{equation}
\Psi_{0}^{N}(\x_{1},\dots,\x_{N})f(t_{N+1})	
\end{equation}	
with $\Psi_{0}^{N}$ the ground state wave function for $N$-particles 
and $f$ as in (\ref{eff}) as a test wave function for the $N+1$-particle 
Hamiltonian. Besides the lemma  one needs the easy to verify fact that the GP 
energy $E^{\rm GP}_{R}$ in the box $\Lambda_{R}$ converges to $E^{\rm GP}$ 
as $R\to\infty$.

The lower bound, and hence the main result, now follows from a 
rigorous version of (\ref{estimate}) in the box $\Lambda_{R}$.  We use 
(\ref{lbd2}) in the boxes $\alpha$; the error here is 
$O(Y_{\al}^{1/17})$ with $Y_{\al}\sim a^3n_{\alpha}/L^3 
\leq a^3N/L^3$. For $Na$ 
fixed, this is $O(N^{-2/17}L^{-3/17})$.
The difference between $\rho_{\al,{\rm max}}$ and 
$\rho_{\al,{\rm min}}$ introduces an additional error, $O(L)$. 
Both error terms are of the same order for $N\to\infty$ if we 
choose $L\sim N^{-1/10}$. We thus obtain
\begin{equation}
E^{\rm QM}(N,a)\geq E^{\rm QM}_{R}(N,a)\geq E^{\rm 
GP}_{R}(N,a)(1-\hbox{(const.\,)}N^{-1/10}).
\end{equation}
The constant may depend on $R$, but this is of no harm, since we 
first take $N\to\infty$ and then $R\to\infty$, using $E^{\rm 
GP}_{R}\to E^{\rm GP}$ and (\ref{scaling}). This proves (\ref{econv}).
Convergence of the densities, (\ref{dconv}), is obtained in a standard
way (see, e.g., \cite{lieb81}) by variation with respect to the
external potential.

\section{Concluding remarks}
\medskip
{\it Bose Einstein 
condensation} in the ground state is a concept that 
involves the full one particle density matrix
\begin{eqnarray}\gamma{}_{N}(\x,\x')
=N\int \Psi_0(\x,\x_2,\dots,\x_N)^*
\Psi_0(\x^\prime,\x_2,\dots,\x_N)d\x_2\cdots d\x_N\end{eqnarray}
and not only its diagonal, $\rho^{\rm QM}(\x)=\gamma{}_{N}(\x,\x)$ 
considered here. Expressed in terms of creation and annihilation 
operators we can also write 
\begin{equation}\gamma{}_{N}(\x,\x')=\langle 
a^*(\x)a(\x^\prime)\rangle_0,
	\end{equation}
with $\langle\cdot\rangle_0$ the expectation value in 
the ground state. As an integral kernel, $\gamma{}_{N}$ is a positive 
trace class operator. If 
$N_{0}$ denotes its largest eigenvalue 
BE condensation means that there is a constant $c>0$ so that
\begin{equation}N_{0}>cN\end{equation}
for all $N$. 
\medskip
This definition applies also if $\langle\cdot\rangle_0$ is replaced by 
a thermal equilibrium state at nonzero temperature.

Note that the density in momentum space is 
\begin{equation}\langle 
\tilde a^*(\p)\tilde 
a(\p)\rangle_0=\int\exp(i\p\cdot(\x-\x^\prime))\gamma{}_{N}(\x,\x^\prime)d\x 
d\x^\prime \  , \end{equation} 
and this differs from $\left|\int 
\exp(i\p\cdot\x)\sqrt{\rho(\x)}d\x\right|^2$ unless $\gamma{}_{N}$ has rank 1.  \medskip

It is often claimed that $\Phi^{\rm GP}$ is (approximately) the 
eigenfunction to the highest eigenvalue of $\gamma{}_{N}(\x,\x^\prime)$ 
and hence that $|\tilde\Phi^{\rm GP}(\vec p)|^2$ gives the momentum 
distribution of the condensate, but this is not proved yet.  In fact, 
so far the only cases with genuine interaction where BE condensation has been 
rigorously established in the ground state are lattice gases at
precisely half filling. (The hard core lattice Bose gas corresponds to
the $XY$ spin $1/2$ model and BE condensation was proved in dimension
$\geq3$ \cite{DLS} and dimension $2$ \cite{LKS1988a}. The hard core
lattice gas with nearest neighbor repulsion corresponds to the
Heisenberg antiferromagnet and condensation was proved for high
dimension in \cite{DLS} and dimension $\geq3$ in \cite{LKS1988}.
Dimension $2$ is still open.)

\end{document}